\documentstyle[aps,prl,floats,graphicx,epsfig]{revtex}   
\twocolumn
\begin{document}
\title{First-order transition in the one-dimensional three-state Potts 
model with long-range interactions }
\author{Zvonko Glumac and Katarina Uzelac}
\address{ Institute of Physics, POB 304,
Bijeni\v{c}ka 46, HR-10000 Zagreb, Croatia}
\maketitle

\begin{abstract}
	The first-order phase transition in the three-state Potts model with 
long-range interactions decaying as $1/r^{1+\sigma}$ has been examined by 
numerical simulations using recently proposed Luijten-Bl\"ote algorithm. 
By applying scaling arguments to the interface free energy, the  Binder's 
fourth-order cumulant, and the specific heat maximum, the change in the 
character of the transition through variation of parameter $\sigma$ was studied.
\end{abstract}

\pacs{PACS numbers: 05.50.+q, 64.60.Cn }

\narrowtext

\section{Introduction}

The critical behavior of models with long-range (LR) interactions  
has been considerably less explored than that of the short-range (SR) ones, 
especially for discrete models where the non-locality of interactions 
makes most of the standard methods ineffective.
These models may, however, exhibit rather complicated  critical 
behavior already in one dimension \cite{DY,KO,CA,GU93}. 
Recent attention has been driven to such models in the
context of studying phenomena related to LR interactions\cite{TS,CM}, 
but also in view of possible equivalence with SR models\cite{TS,LB96}.

 One of the interesting and still not clarified aspects of LR models is the
possible onset of the first-order transition. 
The question naturally arises for the LR Potts model, which in the SR 
interaction case is known to undergo a first-order phase transition 
when the number of states $q$ exceeds a certain limiting value which depends 
on dimensionality, $q_c(d)$\cite{WU,BA,AP}. 

	We have recently pointed out \cite{UG97} that a similar behavior 
indeed can be observed in the $1d$ Potts model with interactions decaying  
with distance as $1/r^{1+\sigma}$ and illustrated it on special cases 
$q=5, \sigma=0.2$ and $q=3, \sigma=0.8$ characteristic of the two regimes.
On the basis of these preliminary Monte Carlo (MC) results on small chains
for $q=3$ and $q=5$, we have concluded on the existence of a $q$-dependent 
threshold value $\sigma_c(q)$ separating the first- and the second-order 
transition regime. 

In the present paper, we focus on a more detailed study of the $1d$ 
three-state Potts model with $0<\sigma<1$.

As is well known, the first-order phase transitions are rather difficult to 
detect and study\cite{JV}. 
Most of standard renormalization group (RG) approaches 
do not distinguish them from the second-order transitions. 

On the other hand Monte Carlo simulations, which in combination 
with finite-size scaling (FSS) represent an efficient tool\cite{BH} to solve 
this problem, are limited when the first-order transition is weak, so that the 
correlation length, although finite, exceeds the considered system size. 
This problem is even more pronounced when dealing with LR interactions, 
where MC simulations become much more time consuming, no matter 
whether Metropolis or different cluster algorithms are used.

Recently, Luijten and Bl\"ote\cite{LB95} have proposed a rather efficient 
algorithm which permits considering quite large sizes in spite of the 
long-range interactions. 
Our intention is to use this approach here in order to perform a more 
systematic analysis of several quantities (interface free energy, Binder's 
fourth order cumulant, specific heat), which serve as criteria for 
distinguishing first- from second-order phase transition.

\section{Model and method}

The model we consider is defined by the Hamiltonian 
\begin{equation}
H = - \sum_{i < j} \; \frac{J}{|i - j|^{1+\sigma}} \; \delta (s_i, s_j) \; ,
\label{eq:hamilt}
\end{equation}
where $J>0$, $s_i$ and $s_j$ denote three-state Potts variables at sites 
$i$ and $j$, respectively, $\delta$ is the Kronecker symbol and summation is 
taken over all pairs of the system.

This model has a phase transition at finite temperature  
for $0<\sigma\le1$ \cite{CA,ACHCHN,GU93}.
No exact results exist for the related critical behavior. 
By analogy with the SR interaction case extended to arbitrary dimension, 
one expects to find a first-order phase transition for low values of 
$\sigma$  and a crossover to a second-order phase transition above a certain 
threshold value $\sigma_c$.
This has been confirmed by our preliminary MC results\cite{UG97} in $1d$ LR 
Potts model which lead to the conclusion, that for $q>2$ the transition is of the 
first-order when $\sigma<\sigma_c(q)$ and of second order above it. 
The threshold $\sigma_c(q)$ is expected to depend on $q$. 
For $q=3$, the $\sigma_c$ was estimated\cite{UG97} to lay above 
$\sigma_{MF}=0.5$, the point separating the mean-field (MF) from the nontrivial 
critical regime in the Ising ($q=2$) case ( where the transition, by symmetry 
reasons (see e.g. ref. \cite{WU} and references therein), remains of the second 
order\cite{AIZFER} in both regimes).
Above the threshold $\sigma_c$, the model undergoes a second-order 
transition with non-classical critical exponents depending on $\sigma$. 
Several renormalization group approaches yield approximate values of these 
exponents, within continuous Ginzburg-Landau functional formalism \cite{PL} 
or in real space \cite{CA,GU93,CM}. 
These latter approaches, however, appear to be insensitive to detect a 
first-order transition.

	In the present approach we shall be using the MC algorithm, recently introduced 
by Luijten and Bl\"ote \cite{LB95}, designed for models with LR interactions.
It is based on Wolf's cluster algorithm \cite{W89} and was applied to several MF 
related problems in the special (Ising) case $q=2$\cite{LB96,LB97}.

	The basic idea of the method is the use of cumulative bond probabilities 
in building the cluster of connected bonds, which drastically reduces the 
number of operations required. 
The results coincide with those obtained with the simple Wolf algorithm, 
but the CPU time can be reduced by several orders of magnitude.\\

The basic quantity in our calculations is the energy probability distribution defined by
\begin{equation}
P_L(E) = \frac{1}{Z_L(K)} {\cal N}_L(E)\ \ e^{-K E}\ , 
\end{equation}
where $K=1/T$ is the inverse temperature,  $J=k_B$,  $Z$ is the partition 
function,  ${\cal{N}}_L(E)$ denotes the number of spin configurations 
corresponding to the energy $E$, and $L$ is the system size.

	Several quantities used for the determination of the temperature driven
first-order transition can be deduced from $P_L(E)$. We concentrate here
on three most important ones: the interface free energy, the Binder's fourth
order cumulant\cite{BIN}, and specific heat. 

The interface free energy is obtained from the shape of the energy 
probabability distribution $P_L(E)$. For temperature driven first-order
transitions, at the transition temperature, $P_L(E)$ has two maxima, 
corresponding to the coexisting ordered and disordered phases. 
The interface free energy is then defined by
\begin{equation}
\Delta F_L  = \ln\left. \frac{P_{LMax}}{P_{Lmin}}\right\vert_{K_L}, \;
\end{equation}
where the finite-chain transition temperature $K_L^{-1}$ has been defined by
requiring that the two maxima are of equal height  $P_{LMax}$.  $P_{Lmin}$ 
denotes the minimum of $P_L(E)$ between them.

The scaling analysis of $\Delta F_L$ can be used to identify the first-order phase 
transition, even in the case of a weak first-order transition, where the correlation 
length, though finite, is large and comparable to the system size\cite{LK0}.
When the transition is of the first-order, $\Delta F_L$ increases with size.
For systems with SR interactions the interface free energy has the dimension 
of surface and scales as $\Delta F_L \sim L^{d-1}$. 
In the present model with LR interactions it is expected to scale as a volume.

	The other two quantities can be derived from the higher energy momenta 
of $P_L(E)$, defined as
\begin{equation}
 \langle E^n \rangle_L = \sum_E  E^n P_L(E). 
\end{equation}

The specific heat $C_L$ (defined per spin in the remaining text) is related to the 
second moment and defined for the system size $L$ by
\begin{equation}
C_L =\frac{K^2}{L^d} \left( \; \langle E^2\rangle_L - \langle E\rangle_L^2 \;
\right
) \; .
\label{e2} 
\end{equation}
According to the FSS theory, in the case of a second-order phase transition 
its maximum scales as 
\begin{equation}
C_{LMax}  \sim L^{\alpha/\nu}, 
\label{c2nd} 
\end{equation}
while for the first-order one it simply scales as volume which in the present 
case means linear dependence on $L$
\begin{equation}
C_{LMax}  \sim L. 
\label{c1st} 
\end{equation}

The calculation of the fourth moment gives Binder's fourth order cumulant
\cite{BIN} defined as
\begin{equation}
V_L^{(4)}=1-U_L^{(4)}/3,
\end{equation}
where
\begin{equation}
U_L^{(4)}  = \; \frac{ < E^4 >_L  }{ < E^2 >_L^2 }  \; .
\label{e4} 
\end{equation}
In the present study, for practical reasons it is easier to deal with $U_L^{(4)}$.
In the thermodynamic limit $U^{(4)}_L$ tends to one, when $K \ne K_c$. 
At $K=K_c$ it still tends to one if the transition is of the second order, while 
it tends to a different constant in the case of a first-order transition.
Together with $\Delta F$, $U^{(4)}$ appears as one of the most 
sensitive criteria for determination of first-order phase transitions 
\cite{BIN,LK1}. 

\subsection{The special case $\sigma = -1 $}

Before proceeding to the presentation of our numerical results, let us summarize the
only analytical results available for the considered model. 
They  can be obtained in the limit $\sigma = -1$ of model (1), where all the 
interactions are of equal strength and the coupling constant is redefined so that  
$K \to K/L$. This is the mean-field limit, which has been extensively studied in 
literature ( see e.g. \cite{WU} ) and solved in the limit $L \rightarrow \infty$ by 
saddle point method\cite{K54}.

The energy $E$ and the entropy $S$ of this model to the leading order in $L$ writes
\begin{equation}
 \frac{E}{L} \; = \; -\frac{1}{2} \sum_{m=0}^{q-1}\left(\frac{L_m}{L}\right)^2 ,
\end{equation}
\begin{equation}
\frac{S}{L} \; = \; - \sum_{m=0}^{q-1} \frac{L_m}{L} \ln \left(\frac{L_m}{L}\right),
\end{equation}
where $L_m$'s, the numbers of particles in the $m$-th Potts state, satisfy  the
condition
$L_0 + \dots + L_{q-1} = L$.
The
transition temperature, the order parameter jump and the latent heat are 
known \cite{K54} for the above model.

To the best of our knowledge, the results for the characteristic quantities
$U_{Max}^{(4)}$ and $C_{Max}$ were not cited in literature.
In order to obtain them, we have used the Lee and Kosterlitz \cite{LK1}
prescription for the energy probability distribution $P_L(E)$ in the vicinity
of the first-order transition temperature $K_c^{-1}$ for large $L$, 
\begin{equation}
 P_L(E) \; = \; \frac{ f\left(Lt\right) \; \delta(E-E_o) + \delta(E-E_{do}) }
               { f(Lt) + 1 },
\label{mfp}
\end{equation}
where $t = (K-K_c)/K_c$, and $E_o$ and $E_{do}$ are the energies of ordered 
and disordered phases, respectively.
The weight function $f(Lt)$ which contains the entire tem\-pe\-ra\-tu\-re  and 
$L$ dependence of $P_L(E)$ is not given explicitly.
Only the limiting behavior of $f(Lt)$, $f(x\to -\infty) \to 0$ and $f(x\to \infty) \to +\infty$, 
is required, in order to have only one $\delta$-function in thermodynamic limit 
for temperatures above or below the transition temperature, while for finite $L$ 
the two $\delta$-functions coexist.
Maximizing $U_L^{(4)}$ and $C_L$ with respect to temperature fixes
the value of the function $f(Lt)$ and leads to
\begin{equation}
 \lim_{L\to\infty} \; U_{L Max}^{(4)} \; =  \; \left[
           1 + \frac{(q-2)^4}{2 \; (q-1) \; (q^2-3q+3)}
           \right]^2 
\label{mfu4}
\end{equation}
and
\begin{equation}
 \lim_{L\to\infty} \; \frac{C_{L Max} }{L} \; = \; \left[
                  \frac{q-2}{2 \; q} \ln (q-1)
                                         \right]^2.
\label{mfc}
\end{equation}

For finite $L$, the exact calculations may be performed numerically for very large
sizes ($L=100\,000$ is easily attained).
They match well with the analytical results (\ref{mfu4}) and (\ref{mfc}) and justify
the conjecture (\ref{mfp}) from which they were derived.

\section{Numerical  Results}

	By using the Luijten-Bl\"ote cluster algorithm, we have been able 
to reach sizes up to $5\, 000$ with reasonable computing time.
The systematic simulations were performed for sizes $L$ ranging from $200$ to 
$3\, 000$, with periodic boundary conditions. The parameter $\sigma$ was 
taken with the increment of $0.1$ in the interval $0<\sigma<1$, where the 
nontrivial transition is to be expected. 
For each set of parameters, $10^6$ clusters were generated. 
Let us also point out that the characteristic temperatures related to the three
quantities considered are different until the thermodynamic limit is reached.
Consequently, an independent numerical effort was needed for localizing the 
characteristic temperatures for each of those quantities for every $L$ and
$\sigma$ considered. 
By combining the Ferrenberg-Swendsen histogram method  \cite{FS} and 
direct calculations, these temperatures have been calculated with numerical 
precision up to the fourth decimal digit in $K_L$, which implies a numerical 
error of approximately 0.1\%. 
We did not go beyond this precision in $K_L$ since the aspect of the calculated
distribution $P_L(E)$ is much rougher than the one obtained with the Metropolis
algorithm, when using a comparable number of steps.
Resulting numerical error in calculated quantities themselves varies from 1-5 \% 
for $P_L(E)$ and $\Delta F_L$ to 1-2\% for $C_{L Max}$ and $U_{L Max}^{(4)}$.

\subsection{Interface free energy}

\begin{figure}[h!!!]
\begin{center}
\leavevmode
\epsfig{file = 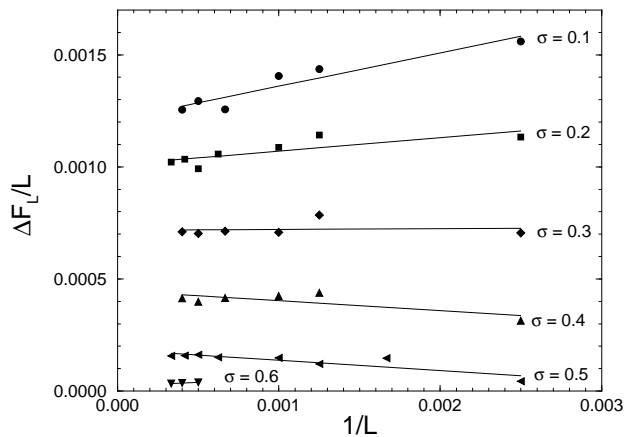, width = 1.0 \linewidth}
\caption[delfsa]{ 
Interface free energy divided by size versus inverse size.
Data for $\sigma = 0.1$ to $0.6$ are included.  Lines are 
obtained by linear regression.
}
\label{delfsa}
\end{center}
\end{figure}
In Fig. 1 are presented the results for the interface free energy divided by size 
versus $1/L$. Only sizes $L\ge 400$ are included.
Points of different shapes correspond to data with different $\sigma$.
It ranges from $0.1$ to $0.6$, where, within considered sizes, the two maxima 
of $P_L(E)$ could be discerned beyond the error limits. 
For  $\sigma = 0.6$ this occurs only when $L\ge 2\, 000$.  

The lines represent the fit of $\Delta F_L / L$ to linear form and illustrate the 
leading correction to the expected scaling form $\Delta F \sim L$, and show 
good agreement with it. 

The weakening of the transition is manifested by the fact that $\Delta F_L/L$
becomes smaller with increasing $\sigma$. This dependence is almost linear 
unless $\sigma$ gets close to the assumed onset of the second order regime.

The other two quantities, $U_L^{(4)}$ and $C_{L Max}$, derived from momenta
(Eqs. (\ref{e2}) and (\ref{e4})), have been calculated in the entire region $0<\sigma<1$.

\subsection{Binder's fourth order cumulant}
 
In order to find out whether in the thermodynamic limit  $U_{LMax}^{(4)}$ will tend to 
one or to a different value, we analyze the maxima of $U_L^{(4)}(K)$ as functions of $L$. 
\begin{figure}[h!!!]
\begin{center}
\leavevmode
\epsfig{file = 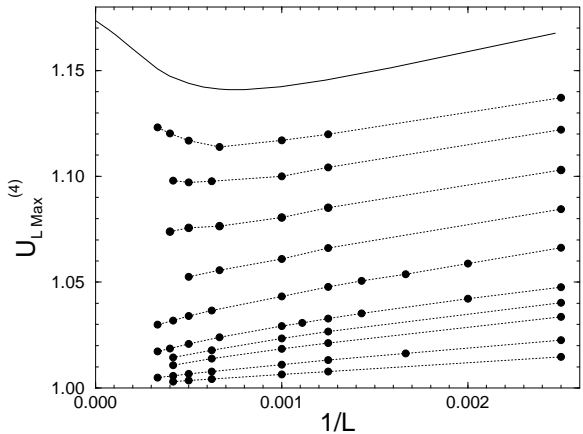,width = 1.0 \linewidth}
\caption[u4l]{ 
Fourth-order cumulant as a function of inverse size.  Points connected with dotted 
lines correspond to $\sigma=0.1,0.2, 0.3,0.4,0.5,0.6,0.65,0.7,0.8,0.9 $, from top to 
bottom. The solid line denotes the results for $\sigma=-1$ limit. 
}
\label{u4l}
\end{center}
\end{figure}
The results for $U_{L Max}^{(4)}$ are summarized in Fig. 2, where the $\sigma=-1$ 
results are also traced for comparison. 
For low values of $\sigma$, $U_{L Max}^{(4)}$ is non-monotonic in $1/L$, similar 
to the $\sigma=-1$ case, so that in the $L \rightarrow \infty$ limit it is clearly 
different from unity.
\begin{figure}[h!!!]
\begin{center}
\leavevmode
\epsfig{file = 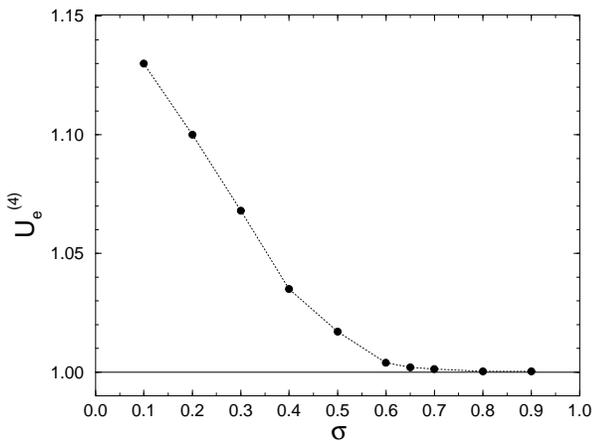,width = 1.0 \linewidth}
\caption[u4e]{  
Extrapolated values of the fourth-order cumulant as a function of $\sigma$.
The constant line at unity is drawn to guide the eye.
}
\label{u4e}
\end{center}
\end{figure}
For higher values of $\sigma$, where the behavior is monotonous, the fit to the 
power-law form $U_{L Max}^{(4)} = U_e^{(4)} + const. \cdot L^x$ was made.
The extrapolated results for $U_e^{(4)}$ are presented on Fig. 3. 
One can observe that the transition between the two regimes occurs in a 
continuous and smooth way with the change of $\sigma$. 
For $\sigma\leq 0.6$ we obtain $U_e^{(4)} \neq 1$ outside the estimated 
error bars (given by the size of the points). 
Also, for $\sigma\geq 0.7$ we conclude that, within the error bars, 
$U_e^{(4)}$ has reached unity.
The additional point $0.65$ can be attributed to both regimes.

These results are consistent with the above ones for $\Delta F$. 

\subsection{Specific heat}

The data for the specific heat maxima  $C_{L Max}$ are summarized in Fig. 4 
in the form of $C_{L Max}/L$ versus $1/L$.
\begin{figure}[h!!!]
\begin{center}
\leavevmode
\epsfig{file = 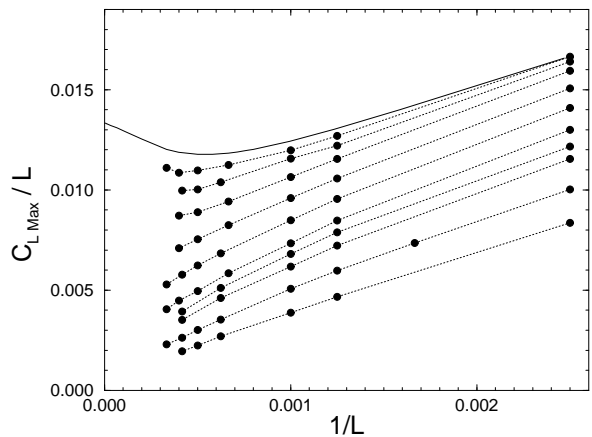,width = 1.0 \linewidth}
\caption[cll]{ 
Maxima of the specific heat divided by size versus inverse size. Solid line denotes 
the case $\sigma=-1$. Points connected with dotted lines correspond to 
$\sigma=0.1,0.2, 0.3,0.4,0.5,0.6,0.65,0.7,0.8,0.9 $, from top to bottom.
}
\label{cll}
\end{center}
\end{figure}
According to Eqs. (\ref{c2nd}) and (\ref{c1st}) and by taking into account that 
$\alpha / \nu $ is much smaller than unity in the region of interest, $C_{LMax}/L$ 
should tend to a constant or to zero value, depending on whether the transition 
is of the first- or second-order, respectively.

Similar as for $U_L^{(4)}$, the curves are non-monotonic and tend to a nonzero 
value in the $\sigma = -1$ case and for low values of $\sigma$. 
For higher values of $\sigma$ the convergence to limit $L \rightarrow \infty$ was 
examined by the fit to the form $C_{L Max}/L = c_0 + const. \cdot L^{-x}$.
The constant $c_0$ is nonzero in the first-order regime, while in the second-order 
one $c_0=0$ and $x=1-\alpha / \nu$. 
This means also that the log-log plot of the curves of Fig. 4 should have a linear 
shape in the latter case, while it will deviate from the straight line in the former case. 
In Fig. 5 log-log plots of $C_{L Max}/L$ versus $1/L$ are given for $\sigma \ge 0.5$.
The curves for $0.5$ and $0.6$ clearly leave the straight line.
\begin{figure}[h!!!]
\begin{center}
\leavevmode
\epsfig{file = 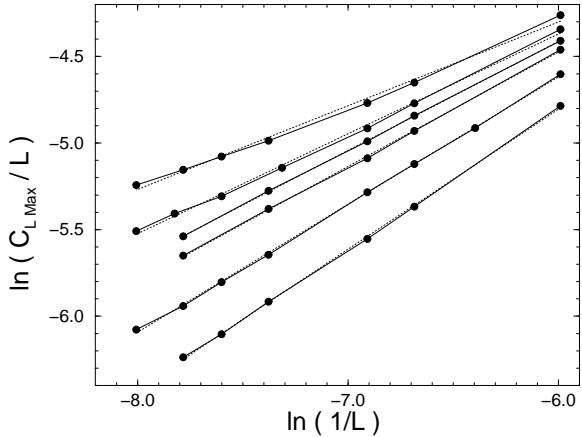,width = 1.0 \linewidth}
\caption[lncll]{ 
Log-log plot of the maxima of the specific heat versus inverse size. The points 
connected by solid lines represent the results for $\sigma = 0.5, 0.6, 0.65, 0.7, 0.8$ 
and $0.9$, from top to bottom. Dotted lines denote the fit to the straight line by linear 
regression.
}
\label{lncll}
\end{center}
\end{figure}
Linear fit is very good for $\sigma \ge 0.65$, indicating the second-order transition. 
The corresponding critical exponent $\alpha / \nu $ following from the fit for 
$\sigma = 0.65, 0.7, 0.8, .0.9 $ is $ 0.36, 0.33, 0.26, 0.19$, respectively, which is 
in poor agreement with values expected from the finite-range scaling (FRS) 
\cite{GU93} calculations, which for the same values of $\sigma$ would give 
$\alpha / \nu = 0.32, 0.27, 0.15, -0.02$.
One should mention, however, that  it is generally a difficult task to extract critical 
exponents with good precision from the spacific heat maxima \cite{HJ34}.

\subsection{Critical temperature}

	The three different characteristic temperatures arise in calculation of 
$\Delta F_L$  and maxima of $U_{L}^{(4)}(K)$ and  $C_{L}(K)$, which all 
should tend to the critical temperature in the thermodynamic limit. 
The $L\rightarrow\infty$ extrapolations of these quantities made by assuming 
power-law corrections in (1/L) are presented in Table I, compared to the existing 
earlier results obtained by   FRS \cite{GU93} and RG \cite{CM}. 
The extrapolation errors are estimated to be in the limits $(\pm 0.01)$ which is 
also comparable to difference between the extrapolations obtained from different 
characteristic temperatures. The agreement with the FRS results is quite good,
discrepancy does not exceed 5 \%, except for $\sigma=0.1$, where the convergence 
of FRS is weakest, and the extrapolation errors in both methods are largest. 
(For $\sigma=0.1$ all the claculations were thus performend up to the sizes
$L=5000$.)
The RG results appear to be systematically larger then ours, and the discrepancy 
with them varies from 7 to 33 \%. Similar discrepancy is obtained for the fit 
to the functional form $K_c \sim \sigma$ in the limit $\sigma \rightarrow 0$, 
conjectured in ref.\cite{TS}.

\begin{table}[[h!!!]
\caption{
Inverse critical temperatures ($K_e(C)$, $K_e(U^{(4)})$, $K_e(\Delta F)$), 
obtained by extrapolation of  $K_L$ compared to FRS extrapolated values 
($K(FRS)$) \protect\cite{GU93} and RG results ($K(CM)$) \protect\cite{CM}.
}
\label{tab1}
\begin{center}
\begin{tabular}{llllll}
 \\
 $\sigma$  & $K_e(C)$ & $K_e(U^{(4)})$ & $K_e(\Delta F)$ &  $K(FRS)$ & $K_e(CM)$
 \\
 \\ \hline
 \\
 0.1   &  0.16   &  0.15  & 0.16  & 0.14  & 0.15  \\
 0.2   &  0.28   &  0.28  & 0.27  & 0.270 &    \\
 0.3   &  0.38   &  0.38  & 0.37  & 0.386 & 0.43  \\
 0.4   &  0.49   &  0.49  & 0.48  & 0.494 &    \\
 0.5   &  0.58   &  0.60  & 0.59  & 0.601 & 0.71  \\
 0.6   &  0.71   &  0.71  & 0.71  & 0.714 &    \\
 0.65  &  0.78   &  0.77  &       & 0.774 &    \\
 0.7   &  0.84  &  0.84   &       & 0.837 & 1.05  \\
 0.8   &  0.98  &  0.99   &       & 0.977 &    \\
 0.9   &  1.14  &  1.13   &       & 1.144 & 1.64  \\
 \\
\end{tabular}
\end{center}
\end{table}

\section{Conclusion}

We have applied MC simulations in combination with FSS in order to examine 
the  onset of the first-order phase transition in the $1d$ three-state Potts 
model with long-range interactions.
 The Luijten-Bl\"ote advanced algorithm for long-range interaction systems 
allowed us to treat successfully considerably large sizes (up to $3\, 000$) 
in a reasonable amount of time, in spite of the long range of interactions. 
The systematic analysis of three quantities: the interface free energy, Binder's 
fourth order cumulant, and specific heat, confirms the existence of two regimes 
in the interval $0<\sigma\leq 1 $.  All three considered quantities give the 
first-order transition for $\sigma \leq 0.6$. 
The transition becomes gradually weaker with increasing $\sigma$ and changes 
smoothly to a second-order transition by continuous variation of $\sigma$. 

For the sizes examined here (up to $3\, 000$) and within the estimated error 
bars, we obtain a second-order transition for $\sigma \geq 0.7$. 

The present approach has, however, limitations to precise determination of the
threshold.
Since the transition close to the threshold becomes arbitrarily weak, the finite 
correlation length, characteristic of the first-order transition, will always be 
much larger than the size of the considered system, close enough to $\sigma_c$. 
Thus, the present result for $\sigma_c$ should be understood only as a lower 
limit  for the possible onset of the second-order phase transition.

For this purpose rather complementary studies should be done, such as the 
one on the dependence of the finite correlation length on $\sigma$.

It is interesting to notice at the end that, according to present results, 
$\sigma_c (q=3)$ falls in the interval between $0.6$ and $0.7$. 
If the correspondence between SR and LR models\cite{LB96} would be extended 
outside the MF regime to the present problem and to the Potts model,  it would 
lead to the conjecture $\sigma_c(q=3)=2/d_c(q=3)_{SR}$ which would give  
$\sigma_c(q=3)$ close to and slightly larger than $0.66$. However, this line of 
argument would also imply that $\sigma_c = 1$ already for $q=4$.


\begin{thebibliography}{...}

\bibitem{DY} F. J. Dyson, Commun. Math. Phys. {\bf 12}, 91 (1969).

\bibitem{KO} J. M. Kosterlitz, Phys. Rev. Lett. {\bf 37}, 1577 (1976).

\bibitem{CA} J. L. Cardy,  J. Phys. A {\bf 14}, 1407 (1981).

\bibitem{ACHCHN} M. Aizenman, J.T. Chayes, L. Chayes, C.M. Newman, 
	    J. Stat. Phys. {\bf 50}, 1, 1988)

\bibitem{GU93} Z. Glumac and K. Uzelac, J. Phys. A {\bf 26}, 5267 (1993).

\bibitem{TS} C Tsallis, Fractals {\bf 5}, 541 (1995).

\bibitem{CM}  S. A. Cannas  and A. C. N. de Magalh\~aes,  J. Phys. A {\bf 30}, 3345
(1997).

\bibitem{LB96}  E. Luijten and H. W. J. Bl\"ote, Phys. Rev. Lett. {\bf 76}, 1557
(1996).

\bibitem{WU}  F. Y. Wu,  Rev. Mod. Phys. {\bf 54}, 235 (1982).

\bibitem{BA} R. J. Baxter, J. Phys. C: Solid State Phys. {\bf 6}, L445 (1973).

\bibitem{AP} A. Aharony and E. Pytte, Phys. Rev. B {\bf 23}, 362 (1981).

\bibitem{UG97} 
K. Uzelac and Z. Glumac, Fizika B {\bf 6}, 133 (1997).

\bibitem{JV}  W. Janke and R. Villanova, Nucl. Phys. B {\bf 489} [FS], 679 (1997);
              W. Janke, in {\it Computer Simulations in Condensed Matter Physics VII},
              eds. D. P. Landau, K. K. Mon and H. B. Sch\"uttler (Springer-Verlag,
Berlin, 1994). 

\bibitem{BH}  K. Binder and H. J. Herrmann, in {\it Monte Carlo Simulation in
Statistical Physics},
eds. M. Cardona, P. Fulde, K. von Klitzing and H.-J. Queisser (Springer-Verlag,
Berlin, 1992).

\bibitem{LB95} E. Luijten and H. W. J. Bl\"ote, Int. J. Mod. Phys. C {\bf 6}, 359
(1995).

\bibitem{AIZFER} M. Aizenman, R. Fern\'andez, Lett. Math. Phys. {bf 16}, 39 (1988)

\bibitem{PL}  R. G. Priest and T. C. Lubensky,   Phys. Rev. B {\bf 13}, 4159
(1976);
              W. K. Theumann and M. A. Gusm\~ ao, Phys. Rev. B {\bf 31}, 379
              (1985).

\bibitem{W89} U. Wolf, Phys. Rev. Lett. {\bf 62}, 361 (1989). 

\bibitem{LB97}
E. Luijten and H. W. J. Bl\"ote,  Phys. Rev. B {\bf 56}, 8945 (1997).

\bibitem{BIN} K. Binder, Phys. Rev. Lett. {\bf 47}, 693 (1981).

\bibitem{LK0}  J. Lee J and J. M. Kosterlitz,  Phys. Rev. Lett. {\bf 65}, 137
(1990).

\bibitem{LK1}  J. Lee J and J. M. Kosterlitz,  Phys. Rev. B {\bf 43}, 3265 (1991).

\bibitem{K54}
T. Kihara, Y. Midzuno and T. Shizume, J. Phys. Soc. Jpn. {\bf 9}, 681 (1954). 

\bibitem{FS}  A. M. Ferrenberg and R. H. Swendsen, Phys. Rev. Lett. {\bf 61}, 2635
(1988).

\bibitem{BS} see e. g. H. W. J. Bl\"ote and R. H. Swendsen, Phys. Rev. Lett. {\bf
43}, 799 (1979)
             and references in \cite{WU}.

\bibitem{HJ34} C. Holm and W. Janke, Phys. Rev. B {\bf 48}, 936 (1993); 
               C. Holm and W. Janke, J. Phys. A {\bf 27}, 2553 (1994). 
 
\end{thebibliography}
\end{document}